# An Analysis of the Accuracy of the Added Length De-Embedding Methods for Coaxial to Waveguide Adapters in the X-Band

Eric Bauman [#1], James DePoy [#2], Sterling Light [#3], Arya Menon[#4], Marco Iskander[#5]
[#] Texas A&M University, USA
{[1] ericbauman, [2] jamesdepoy, [3] s.a.light, [4] aryamenon, [5] marcoi} @tamu.edu

***Abstract*—This paper assesses the accuracy of the "added length method" (also known as *port extension* or *channel offset*) for de-embedding coaxial-to-waveguide transitions in the X-band, comparing it to full waveguide calibration on both Keysight and Rohde & Schwarz VNAs. The study examines the trade-offs between cost and measurement accuracy, highlighting the feasibility of using these mathematical correction techniques in the absence of dedicated waveguide calibration kits or full S-parameter characterizations of the transitions. The analysis concluded that when using coaxial-to-waveguide transitions the added length method produces accurate results for transmission measurements only.**

## I. Introduction

Modern Vector Network Analyzers (VNAs) have many different de-embedding techniques for use in a variety of testing scenarios where a device under test (DUT) cannot be measured directly by a coaxial cable [1].

One of those scenarios is measuring waveguide DUTs, which requires adapters when connecting to coaxial cables. Waveguide calibration kits are costly and specific to the type of waveguide being used, such that they are often unavailable. Another commonly used approach is to perform a 2-port calibration using coaxial calibration kits and then de-embed the coaxial-to-waveguide transitions using S-parameter models by importing a touchstone file into the VNA [2] - obtained through EM simulations [3], through measurements [4] or through iterative methods [5] under the assumption that the two transitions are identical. The former is often challenging since the geometrical information of commercial waveguide to coaxial adapters and internal tuning elements are rarely available in datasheets, making it challenging to accurately model. This paper will refer to the aforementioned category of de-embedding as "*direct de-embedding*". However, modern VNAs also offer a variety of simpler correction approaches within their firmware in the absence of a full S-parameter characterization of the adapters, using techniques to mathematically extend the reference plane towards the DUT. *"Port extension*" [7] by Keysight and "*channel offset*" by Rohde and Schwarz (R&S) [8] are examples of these techniques. This paper will call this category of de-embedding the "*added length method*". These techniques "mathematically [extend] the measurement plane towards the DUT" [1] by assuming a length of transmission line and de-embedding it without the need to import a touchstone file to model the coax-to-waveguide transition. These techniques are widely used in on-chip measurements using probes [6]. However, to the best of the authors' knowledge, there is little research on the validity of this method for waveguide DUTs. This paper will assess the accuracy of these techniques on waveguide DUTs in the X-band on both Keysight and R&S VNAs as compared to a full waveguide calibration. The paper will examine the trade-offs of using these techniques versus the control case using a waveguide calibration kit.

## II. Definition of Terms

A definition of terms is necessary to distinguish the various ways that VNAs perform the added length method, since different terminology is used by different manufacturers. The added length method is known as "*port extension*" [7] to Keysight and "*channel offset*" [8] to R&S.

Within Keysight's *port extension* and R&S's *channel offset* there are several variations. The main split is between manual techniques that require inputting the length and the loss of the fixture, and automatic techniques that require measuring a termination, usually either open or short terminations.

- The manual techniques will be denoted with the word "Manual", either as "*manual port extension*" for Keysight or "*channel offset, manual length input*" for R&S.
- Keysight's automatic reference plane extension algorithm will be referred to as "*auto port extension*", with the termination(s) used to measure it placed afterwards with a comma, e.g. "*Auto port extension, short*".
- R&S's *channel offset* has two different automatic algorithms. The R&S equivalent to Keysight's *auto port extension* is known as "*Fixture compensator*", and will be referred to in the same way, i.e. "*Fixture compensator, short*".
- R&S VNA has other techniques that do not require standards and are accessed by selecting the option for either "*auto length*" or "*auto length and loss*" [8] and thus will be referred to as "*channel offset, auto length*" and "*channel offset, auto length and loss*".

## III. Procedure

In the following sections, all experiments were conducted using X-band waveguides. Two angled SMA-to-WR90 adapters with a physical length of 26.41 mm from the back wall to the flange of the waveguide were used to connect to an additional WR-90 waveguide section of length 34.60 mm. The vector network analyzers used were a Keysight PNA-X N5424B (Firmware Version A.15.71.11) and a Rohde and Schwarz ZVA 67 (Firmware Version 4.11).

Both 1 and 2 port measurements were taken. The DUT for the one port measurement was a matched termination, a Lab-Volt model 9531-00, and the DUT for the two-port measurement was a 6 dB attenuator, a Lab-Volt model 9533-00.

The control measurements for all DUTs were performed on the PNA-X with a Keysight X11644A X-band waveguide calibration kit. For all other measurements, the VNA was calibrated to the end of the cables prior to the waveguide adapter using a Keysight 85052B 3.5mm coaxial cable calibration kit. The VNA was calibrated over a frequency range of 8 to 12 GHz with a frequency step of 1 MHz.

On the PNA-X, *port extension* was thoroughly tested. *Port extension* has multiple options, which are *manual length input* and *automatic port extension* by measuring short, open, or both. For the manual input, the advanced settings were used, as this allowed the medium to be set to *waveguide*. The cutoff frequency was set to *6.557 GHz*, based on WR90 theoretical cut-off. The physical length was set to the combined length of the extra waveguide section and the depth of the adapter (*61.01mm*), and the loss was set to *0.159 dB at 8 GHz* and *0.0858 dB at 12 GHz*, which was half the measured insertion loss of the adapters and waveguide sections when placed back-to-back. While setting the medium to *waveguide* is not available with the *auto port extension* [7], its accuracy will still be assessed in this work. For *automatic port extension*, the necessary termination(s) were connected to each port and the measurement was selected and performed on each port individually. For the short termination, a short plate was used as shown in Fig. 1, and for the open termination, a sliding short set to 9.94mm was used, which is a quarter wavelength ($\lambda_g/4$ for $TE_{10}$) at 10 GHz for a WR90 waveguide as shown in Fig. 2.

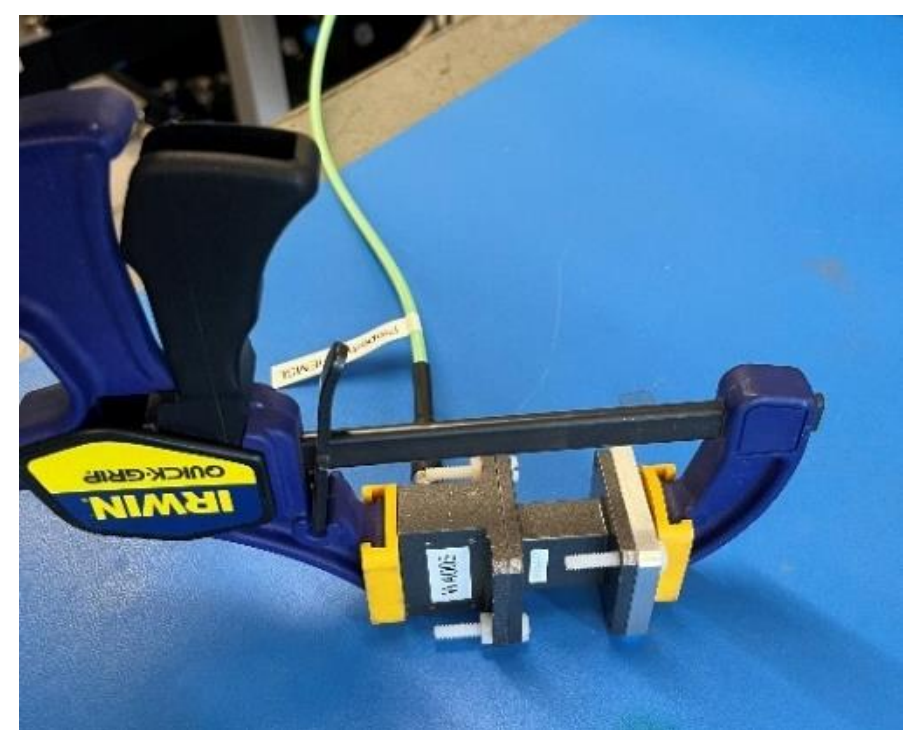

Fig. 1. Short termination plate used for measuring *auto port extension* & *fixture compensator*. A clamp was used to ensure a tight seal.

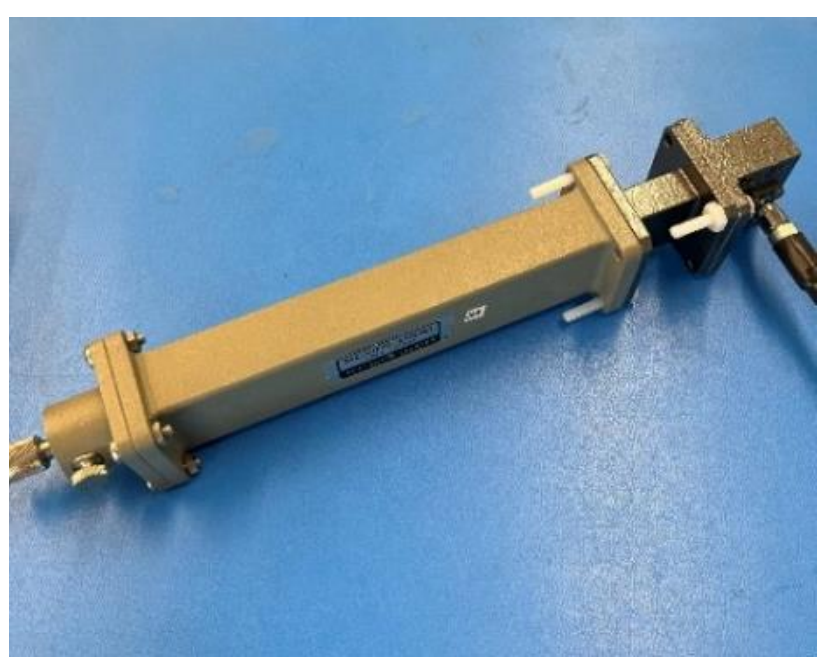

Fig. 2. The sliding short used for modeling the open circuit needed for the *auto port extension* and fixture compensator setup.

On the ZVA 67, the techniques tested were the *channel offset* techniques, including the *fixture compensator*. *Channel offset* not only has a manual length input option, for which the same parameters were used as on the PNA-X, but it also has the "*auto length*" and "*auto length and loss*" options, which are both a single button selection. For *fixture compensator*, the same terminations were used as for the *auto port extension*, measured for each port individually.

## IV. Results and Analysis

### A. *1-port DUT Measurements (matched termination, Lab-Volt 9531-00)*

Fig. 3 shows the S11 magnitude and phase plots for the different added length methods on Keysight and R&S VNAs in comparison with the control case (using waveguide calibration kit).

### B. *2-port DUT Measurements (6-dB Attenuator, Lab-Volt 9533-00)*

Fig. 4-7 show the S-parameters' magnitudes and phase plots for the different added length methods on Keysight and R&S VNAs in comparison with the control case (using the waveguide calibration kit).

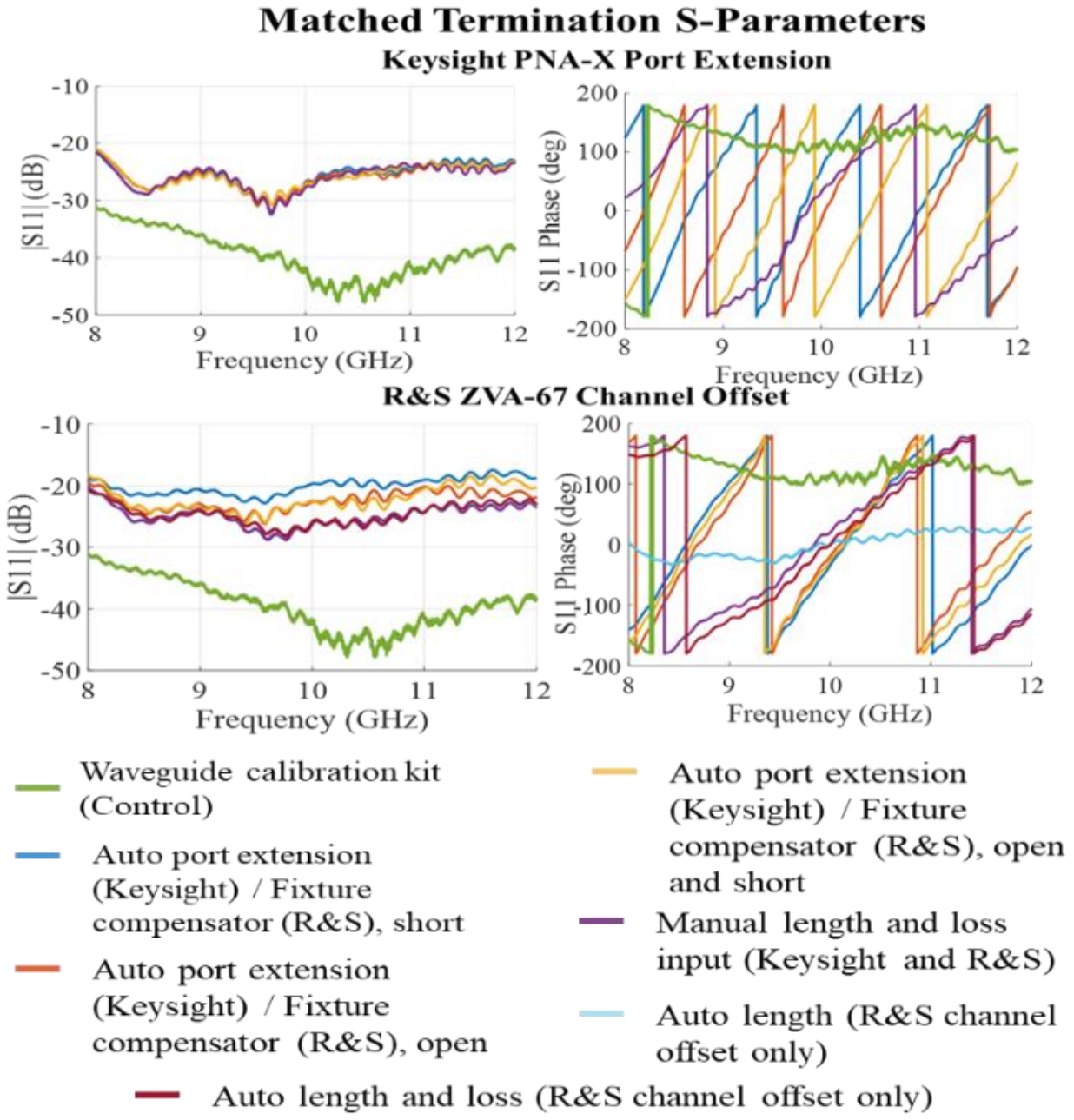

Fig. 3. S-parameter measurement for matched termination DUT for the different added length methods on both Keysight and R&S VNAs.

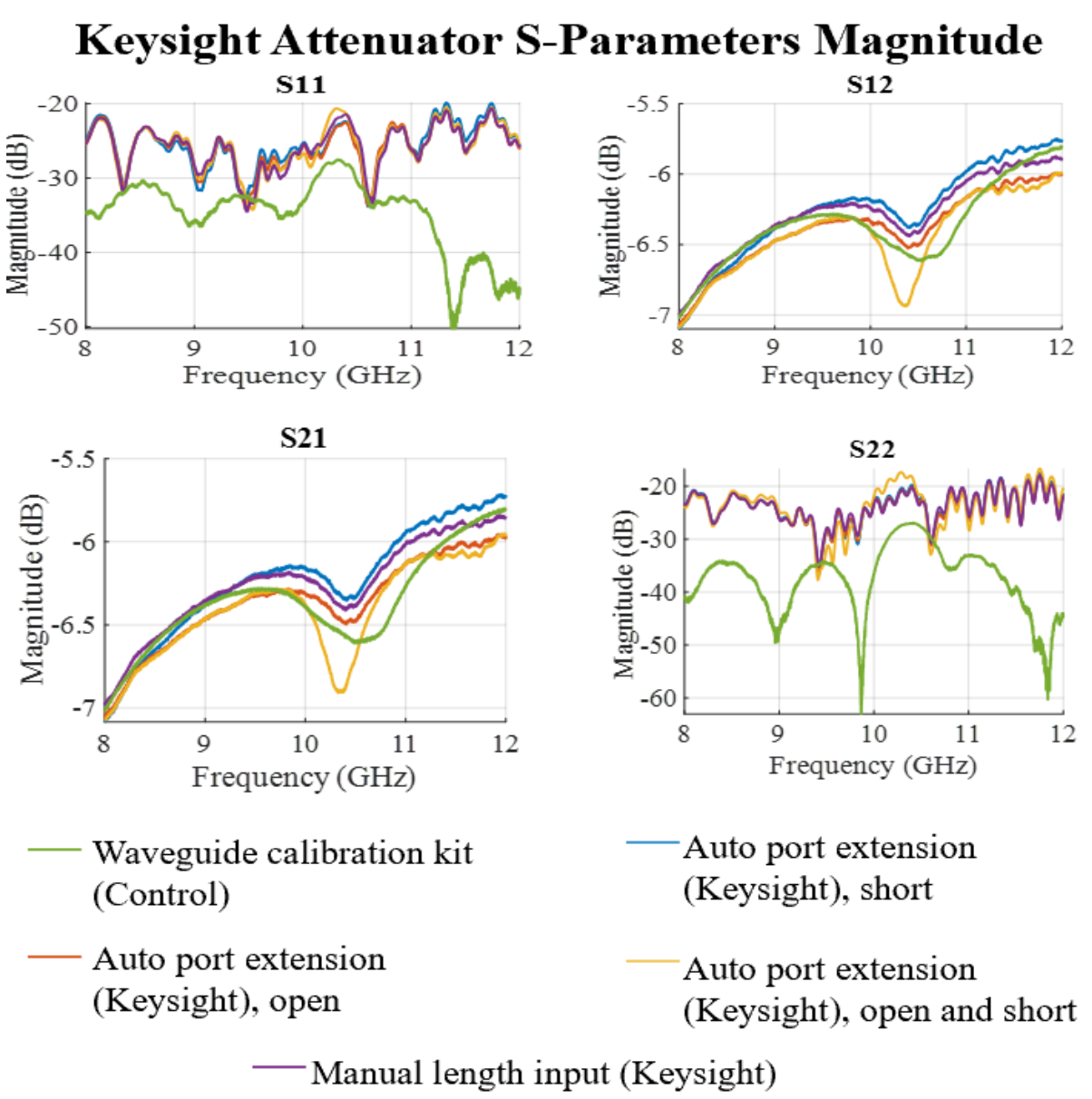

Fig. 4. Magnitude of S-Parameters for the 6dB attenuator DUT measured using *Keysight port extension*.

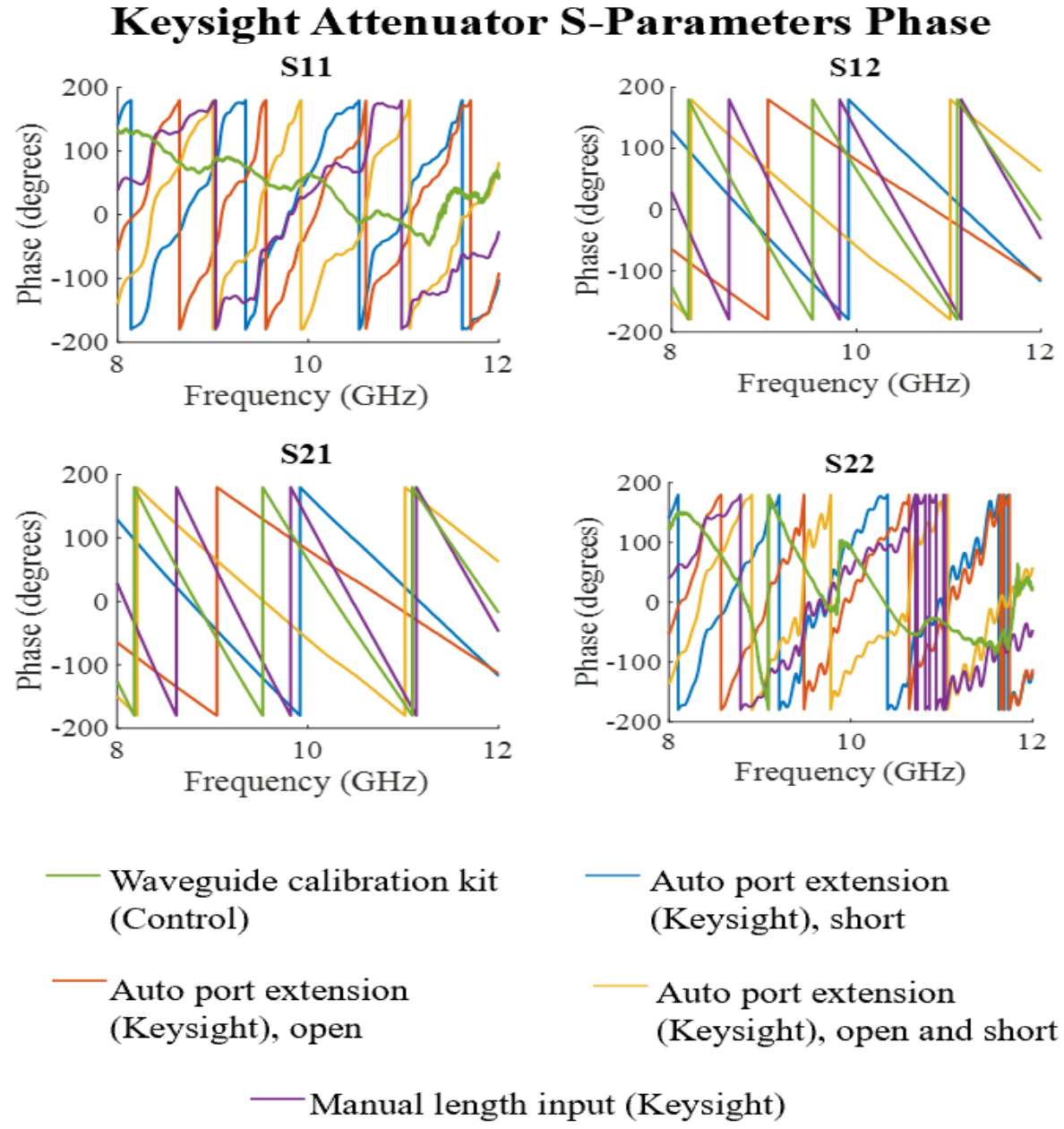

Fig. 5. Phase of S-Parameters for the 6dB attenuator DUT measured using *Keysight port extension*.

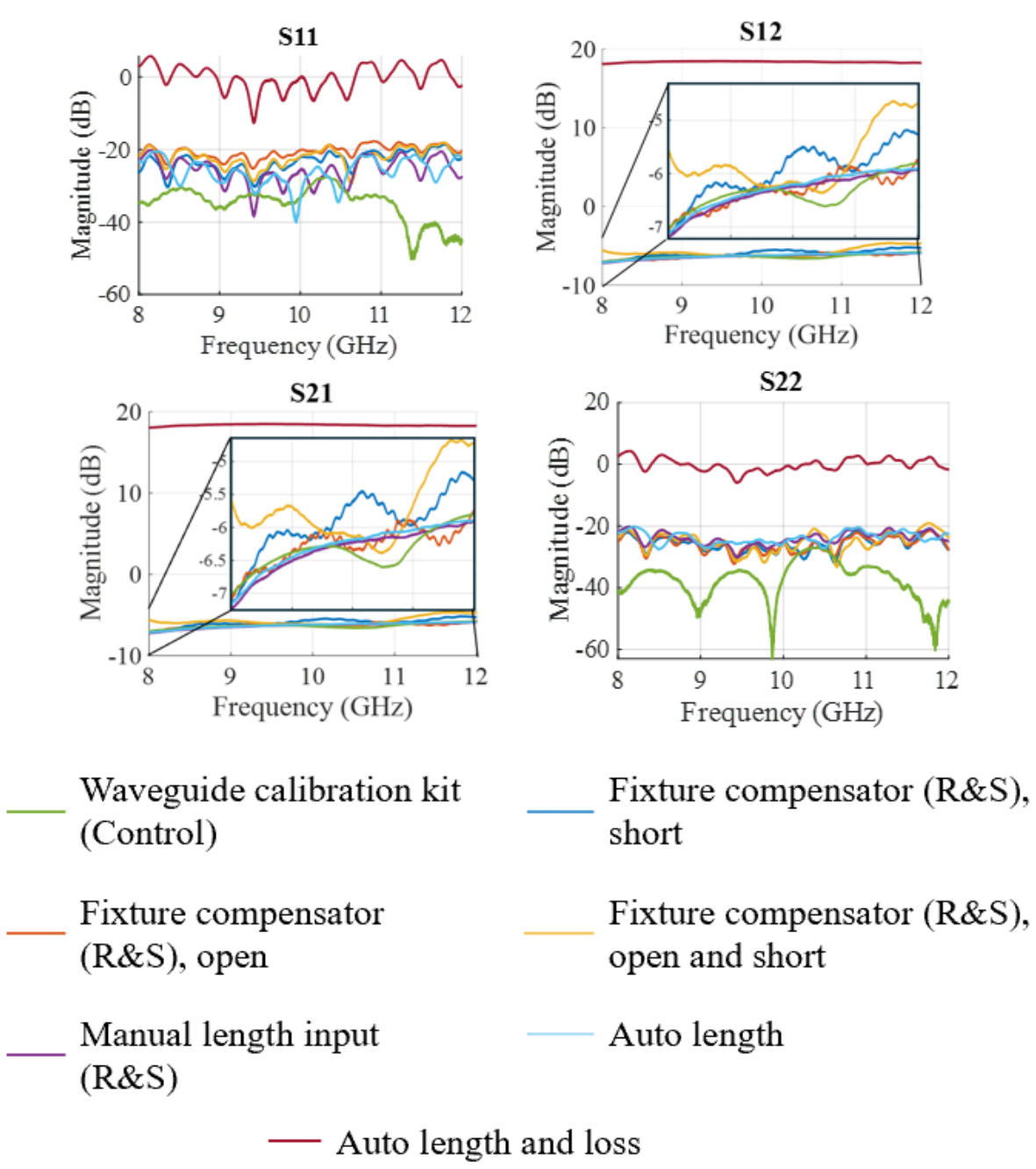

Fig. 6. Magnitude of S-Parameters for the 6dB attenuator DUT measured using *R&S channel offset*.

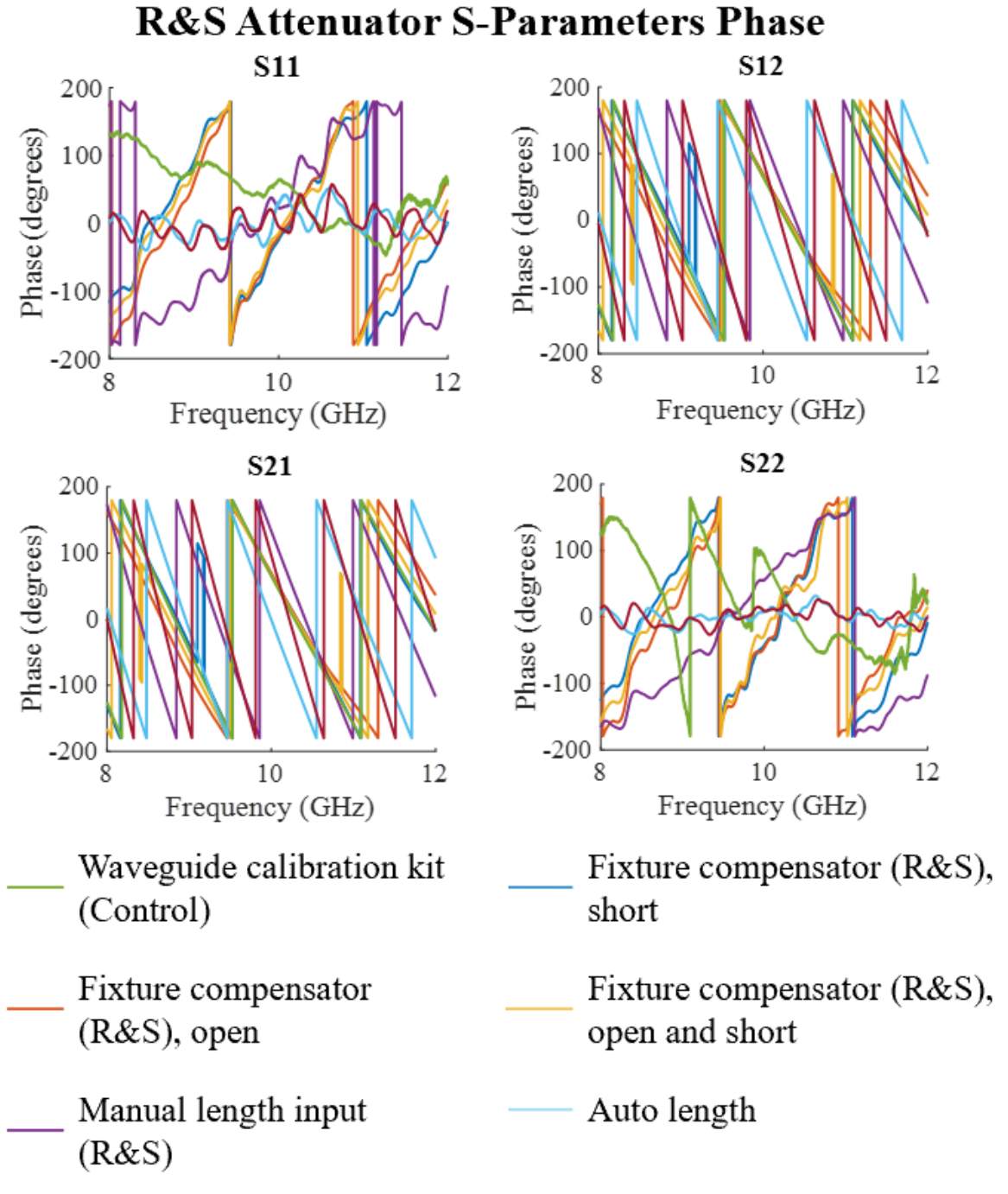

Fig. 7. Phase of S-Parameters for the 6dB attenuator DUT measured using *R&S channel offset.*

## C. Relative Error

The percentage error in the magnitude (without considering phase error) in reference to the control case was calculated using the following formula:

$$\text{Magnitude Error} = \frac{|S_{ij}^{A}| - |S_{ij}^{B}|}{|S_{ij}^{B}|}. \quad (1)$$

where, $S_{ij}^{A}$ is the measured S-parameter and $S_{ij}^{B}$ is the S-parameter in the control case. As evident by Fig. 3, 4 and 6 the reflection measurements deviate significantly from the control case. However, the transmission measurements (S21) are comparable to the control case. Hence, Fig. 8 shows the relative error in magnitude for the different added length methods for the transmission case (S21).

The total relative error was also calculated using the following formula, which captures both magnitude and phase errors:

$$\text{Relative Total Error} = \frac{|S_{ij}^{A} - S_{ij}^{B}|}{|S_{ij}^{B}|}. \quad (2)$$

This equation is adapted from [9] and modified so that the error is normalized to only the control case. Since it takes the complex difference of the two S-parameters, it is a combined measure of the magnitude and phase errors. The relative total error for the transmission case is included in Fig. 9.

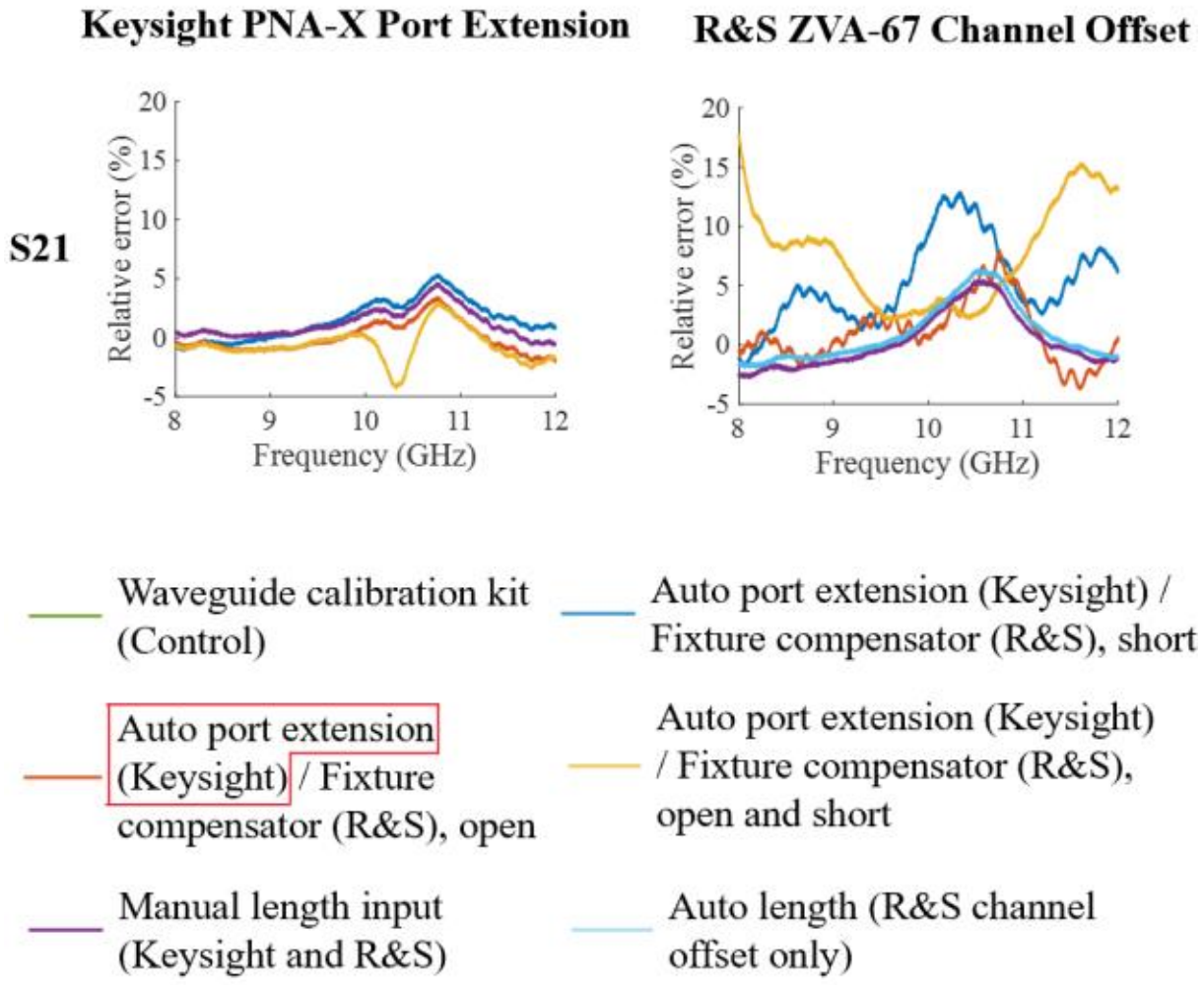

Fig. 8. Relative error in magnitude for the different added length methods on both Keysight and R&S VNAs for the 6dB attenuator DUT, excluding *R&S channel offset, auto length and loss*. Boxed: the method with the least error.

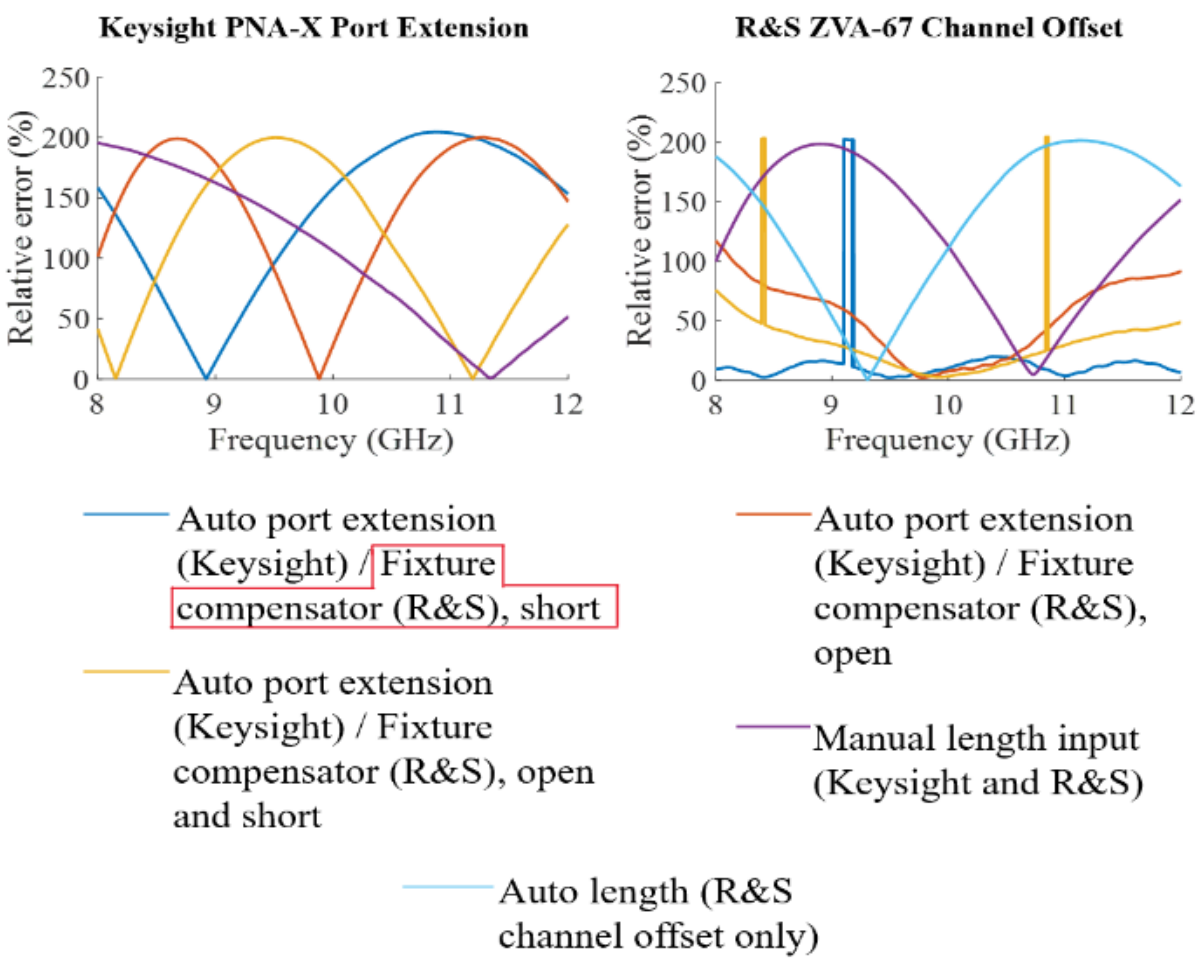

Fig. 9. Relative total error in magnitude and phase of the different added length methods on both Keysight and R&S VNAs for the 6dB attenuator DUT, excluding *R&S channel offset, auto length and loss.* Boxed: the method with the least error.

# V. Discussion

For the matched termination case, Fig. 3 displays that S11 measurements taken using added length methods have magnitudes that differ significantly (5-20 dB) from measurements taken using the waveguide calibration kit. The errors in the measurements of S11 are above or equal to 80% of the magnitude of the average between the measurement and the control case. This means that the error is always either nearly as large or larger than the measurement itself.

Regarding the 6dB attenuator DUT, Fig. 4 and 6 show that the magnitudes of the reflection measurements (S11 and S22) differ significantly (5-20 dB) from the control case similar to the 1 port DUT case. However, the transmission measurements (S12 and S21) display more accuracy in magnitude (~ 0.5 dB) across measurement options, except for the *R&S auto length and loss*. *Keysight port extension* methods are within +/-5% and *R&S channel offset* methods are within +/-15% for magnitude of S21. When it comes to phase error for the transmission measurements, the *R&S fixture compensator* best matches the control case. Note that a spike is observed at the point where the phase is wrapped. For all other methods, the phase error is large due to improper phase corrections. The *R&S fixture compensator*, *short* case produces the best results in this study for transmission measurement considering the magnitude and phase.

Ultimately, it is apparent that reflection measurements taken using added length methods significantly differ from those taken with the waveguide calibration kit, while transmission measurements agree no matter the method used, excepting *R&S channel offset*, *auto length and loss*. The phases measured using added length methods have different slopes over frequency than those measured using the calibration kit, excluding the phase measured with the *R&S fixture compensator*.

The cause of these errors likely lies in the invalidity of assumptions that are made when using the added length methods in the case of waveguides. Although Keysight and R&S VNA user manuals do not disclose the exact mathematical correction formulas used for *port extension* and *channel offset*, the fundamental assumption of the added length method is that constant extra electrical length of perfectly matched transmission line is added between the calibration plane and the DUT with negligible additional losses being introduced [8]. Additional assumptions are based on the method used. These assumptions pertain to aspects such as delay, loss, velocity, and media type. Manual methods permit the user to manually configure each of these assumptions, whereas automatic methods will use either ideal transmission line characteristics or measured standards. However, when using a waveguide, the fundamental assumption is violated. In a waveguide, the phase constant $\beta$ varies with frequency, causing the electrical length to change over the operational range. Additionally, the characteristic impedance varies with respect to frequency, causing a non-trivial impedance discontinuity when changing mediums between coaxial cable and waveguide. The transmission mode also changes from a dominant TEM mode in a coaxial cable to a dominant TE10 mode in a WR-90 waveguide, which causes disruption in the propagating modes that is most likely not fully captured in the corrections. Furthermore, the coaxial-to-waveguide adapter introduces additional discontinuities and reflections which results in the high discrepancy in the reflection measurements. Table I summarizes the different options for characterizing waveguide DUTs and the conclusions.

TABLE I.
OPTIONS FOR CHARACTERIZING WAVEGUIDE DUT ON VNAS

| | **Cost** | **Speed** | **Accuracy** |
|---|---|---|---|
| Using waveguide calibration kit | High | Fast | High |
| Direct de-embedding | Average | Slow | Moderate |
| Added length methods | Lowest | Fast | **Moderate for magnitude of transmission**<br>**Poor for reflection** |

## VI. CONCLUSION

This paper has assessed the accuracy of various added length methods available in commercial VNAs when used for de-embedding coaxial-to-waveguide adapters. It is faster than obtaining a full S-parameter characterization of the adapters (or direct de-embedding) and much cheaper than obtaining a waveguide calibration kit.

For reflection measurements, it has consistently high errors in both magnitude and phase, no matter what instance of the method is used. However, for transmission measurements, the magnitude stays within 10% tolerance using *Keysight port extension* and most *R&S channel offset* techniques and within 20% tolerance using *R&S fixture compensator*. When both magnitude and phase of the transmission measurements are considered, *R&S fixture compensator, short* is most accurate.

## ACKNOWLEDGMENT

The authors would like to thank the Intelligent Electromagnetic Sensor Laboratories (iEMSL) at Texas A&M University for providing access to facilities and resources that supported this research.

## REFERENCES

[1] Keysight Technologies, 5980-2784EN. (2024). *Application Note: De-Embedding and Embedding S-Parameter Measurements Using a Vector Network Analyzer* [Online]. Available: https://www.keysight.com/us/en/assets/7018-06806/application-notes/5980-2784.pdf

[2] Simion, S. (2022). A Simple Method for De-embedding the Coaxial to Waveguide Transitions. In 2022 International Semiconductor Conference (CAS) (pp. 77-80). IEEE.

[3] Eugene, Mayevskyi and Fabrizio, Zanella. "SPICE modeling of a backplane to daughtercard link from an EM simulation environment," *International Engineering Consortium*, vol. 2, pp. 1069 – 1082, 2007. [Abstract]. Available: Engineering Village, https://www-engineeringvillage-com.srv-proxy2.library.tamu.edu/app/doc/?docid=cpx_6e3d6013a225f0857M6dda2061377553. [Accessed: Feb. 7, 2025].

[4] S.Z., Ibrahim, M.S., Razalli, W.F., Hoon, and M.N.A., Karim, "Six- Port interferometer for direction-of-arrival detection system," In *IEEE International Symposium on Systems Engineering (ISSE),* 2016, IEEE. doi: 10.1109/SysEng.2016.7753161

[5] Glock, H.W. and van Rienen, U, "An iterative algorithm to evaluate multimodal S-parameter measurements," *IEEE Transactions on Magnetics,* vol. 36, no. 4, pp. 1841 – 1845, August 2002. [Abstract]. Available: IEEEXplore, https://ieeexplore.ieee.org/document/877803

[6] Amakawa, S., Katayama, K., Takano, K. Yoshida, T., and Fujishima, M., "Comparative analysis of on-chip transmission line de-embedding techniques," In *2015 IEEE International Symposium on Radio-Frequency Integration Technology (RFIT),* 2015, IEEE. doi:10.1109/RFIT.2015.7377897

[7] Keysight, *Keysight PNA Series Network Analyzers.* Keysight, 2019.

[8] Rohde and Schwarz, *R&S®ZVA / R&S®ZVB / R&S®ZVT Operating Manual (Version 33),* 2020.

[9] M. Resso, E. Bogatin and A. Vatsyayan, "A new method to verify the accuracy of de-embedding algorithms," *2016 IEEE MTT-S Latin America Microwave Conference (LAMC)*, Puerto Vallarta, Mexico, 2016, pp. 1-4, doi: 10.1109/LAMC.2016.7851299.